\documentclass[reprint,superscriptaddress,nofootinbib,amsmath,amssymb,aps,prl,notitlepage]{revtex4-1}

\usepackage[a4paper, margin=2.50cm]{geometry}
\usepackage[dvipsnames]{xcolor}
\usepackage{bbm}
\usepackage{textcomp}
\usepackage{tcolorbox}
\tcbuselibrary{breakable}
\usepackage{graphicx}
\usepackage{bbm}
\usepackage{fancyhdr, color}
\usepackage{marginnote}
\usepackage[colorlinks]{hyperref} 
\hypersetup{
    citecolor = blue,
    linkcolor = black
}

\definecolor{yellow}{HTML}{FFDB25}
\definecolor{light_yellow}{HTML}{FFF6B3}
\definecolor{mag}{HTML}{D33EBE}
\definecolor{red}{HTML}{FF006C}
\definecolor{blue}{HTML}{0046FF}
\definecolor{az}{HTML}{00B9E6}

\usepackage[framemethod=tikz]{mdframed}
\usepackage{framed}
\newcommand{\tr}{\mbox{tr}}

\newcommand{\com}[1]{\textbf{\color{OliveGreen}[[#1]]}}

\newcommand{\diag}{\textrm{diag}}

\def\bra#1{\langle{#1}|}
\def\ket#1{|{#1}\rangle}
\def\braket#1{\langle{#1}\rangle}
\newcommand{\ketbra}[2]{\ket{#1}\!\bra{#2}}

\def\BraVert{\egroup\,\mid\,\bgroup}

\newcommand{\chqwcorr}{{\color[rgb]{0.4,0.2,0.9} \sc Qbook:Ch.32}} 

\begin{document}
\fontfamily{ppl}\selectfont


\title{\fontfamily{phv}\selectfont \textbf{Quantum Batteries}}

\author{Francesco Campaioli}
\email{francesco.campaioli@monash.edu} 
\affiliation{School of Physics and Astronomy, Monash University, Victoria 3800, Australia}
\author{Felix A. Pollock}
\affiliation{School of Physics and Astronomy, Monash University, Victoria 3800, Australia}
\author{Sai Vinjanampathy}
\email{sai@phy.iitb.ac.in}
\affiliation{Department of Physics, Indian Institute of Technology Bombay, Mumbai 400076, India}
\affiliation{Centre for Quantum Technologies, National University of Singapore, 3 Science Drive 2, Singapore 117543.}
\begin{abstract}
{\noindent This chapter is a survey of the published literature on quantum batteries -- ensembles of non-degenerate quantum systems on which energy can be deposited, and from which work can be extracted. A pedagogical approach is used to familiarize the reader with the main results obtained in this field, starting from simple examples and proceeding with in-depth analysis. An outlook for the field and future developments are discussed at the end of the chapter.}
\end{abstract}


\date{\today}

\maketitle

\tableofcontents

\thispagestyle{empty}

\section{Introduction} \label{sec:intro} 
\noindent
The thermodynamic role of batteries is that of work reservoir: In practice, classical
batteries are electrochemical devices that store energy supplied by external sources and provide power to other machines, allowing for their remote usage. Their pervasive presence in our daily life has turned them into indispensable components, whose size and storage capability range from large 500kWh vehicle traction batteries to tiny 100mWh miniature cells used for implanted medical devices and calculators \cite{Vincent1997}. Along with the constant miniaturization of such user devices, batteries are also required to be smaller and smaller, thus, as their unit cells approach the size of molecules and atoms, their description has to account for quantum mechanical effects \cite{Goold2016,Vinjanampathy2016a,Campaioli2017}.

This premise leads to the study of 
\emph{quantum batteries} (QBs), which have been introduced by R. Alicki and M. Fannes in 2013 as small quantum mechanical systems that are used to temporarily store energy, in order for it to be transferred from production to consumption centers \cite{Alicki2013}. Further works have used the same formal definition, describing QBs as $d$-dimensional quantum systems with non-degenerate energy levels from which work can be reversibly extracted -- and on which energy can be reversibly deposited -- by means of cyclic unitary operations. It is thus easy to see the relationship of the battery charging problem with the emerging area of quantum thermodynamics \cite{Hovhannisyan2013}.

An immediate consequence of the quantum thermodynamic context is the use of notions and techniques of quantum information, which has brought interesting conclusions and insights. For instance, entanglement and other quantum correlations (see also chapter \chqwcorr) have been addressed in refs. \cite{Alicki2013,Hovhannisyan2013} as possible resources to improve work extraction tasks. The limits to charging power have been set in \cite{Binder2015,Campaioli2017} recalling the quantum speed limit of unitary evolution \cite{Deffner2017}, and using other tools of quantum optimal control theory \cite{Werschnik2007,Caneva2009,Brouzos2015}. However, this blossoming research field has to address many other questions, such as the stabilization of stored energy and the practical implementation of quantum batteries, offering a vast research panorama on both theoretical and experimental ends \cite{Campaioli2017,Ferraro2018b,Friis2017a,Le2018}. 

This chapter will review the main results in this area with simple examples. We will assume the reader is familiar with standard techniques in quantum thermodynamics.
The first section is dedicated to the task of work extraction, introducing the concept of passive states -- from which energy can no longer be reversibly extracted via unitary cycles -- and exploring the limits of classical and quantum extraction protocols. The second section treats the charging of quantum batteries, demonstrating the advantage that entangling operations have over their classical counterparts, while considering the limits of physically realizable charging schemes. In the third section we review some possible implementations of quantum batteries, concluding with a general discussion. 

\section{Work extraction}
\noindent
The study of work extraction from small quantum systems -- regarded as batteries -- via reversible cyclic operations starts with the aim of defining the thermodynamical bounds and principles that are valid at those scales where a quantum mechanical description becomes necessary \cite{Alicki2013}. The intention is to look at the limits of extractable work allowed by quantum mechanics and compare them to their classical counterpart, while looking for possible advantages.

\begin{figure}
    \centering
    \includegraphics[width = 0.30\textwidth]{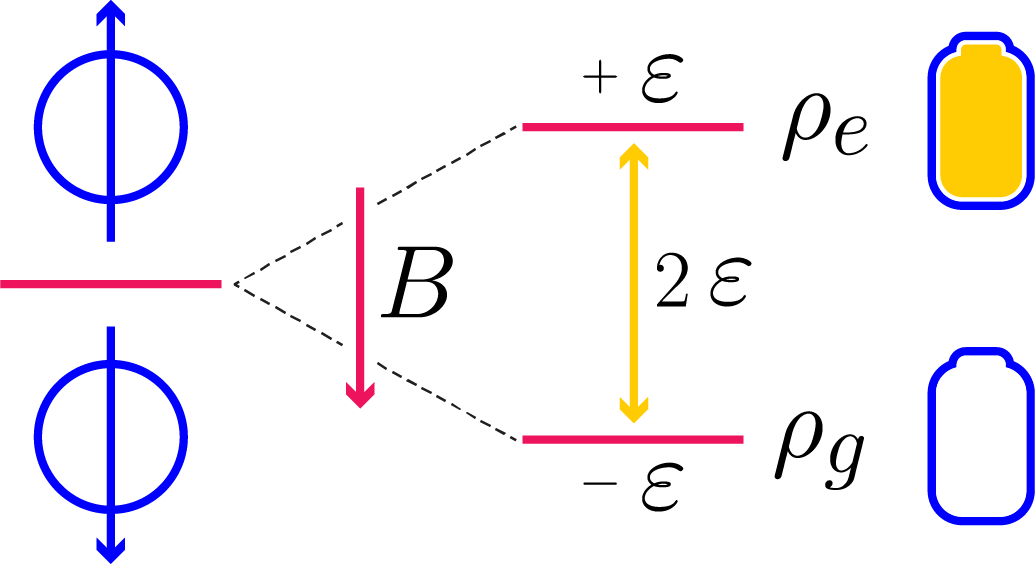}
    \caption{A simple quantum battery could consist of a spin-{\textonehalf} system immersed in a uniform magnetic field $B$, whose internal Hamiltonian 
has energy levels $\mp\varepsilon$, associated with the eigenstates $\ket{\mp 1}$, respectively. A pure state $\rho_e=\ketbra{1}{1}$ ($\rho_g=\ketbra{-1}{-1}$) is considered a charged (empty) battery, since no work can be deposited onto (extracted from) it, with respect to the internal Hamiltonian $H_0$.}
    \label{fig:2-levle-battery}
\end{figure}

Let us start from a single \emph{closed}\footnote{In thermodynamics a \emph{closed} system is only allowed to exchange either work or heat, in contrast with an \emph{isolated} system which is not allowed to exchange either of them. Here, by closed, we mean isolated quantum system undergoing Schr\"odinger evolution, but whose initial state can be mixed.} quantum battery, the fundamental unit of this discussion. It consists of a $d$-dimensional system with associated internal Hamiltonian $H_0$
\begin{equation}
    \label{eq:internal_hamiltonian}
    H_0=\sum_{j=1}^d\varepsilon_j\ketbra{j}{j}, 
\end{equation}
with non-degenerate energy levels $\varepsilon_j<\varepsilon_{j+1}$. An example of a 2-level quantum battery is given in fig. \ref{fig:2-levle-battery}.

\begin{figure}
    \centering
    \includegraphics[width = 0.45\textwidth]{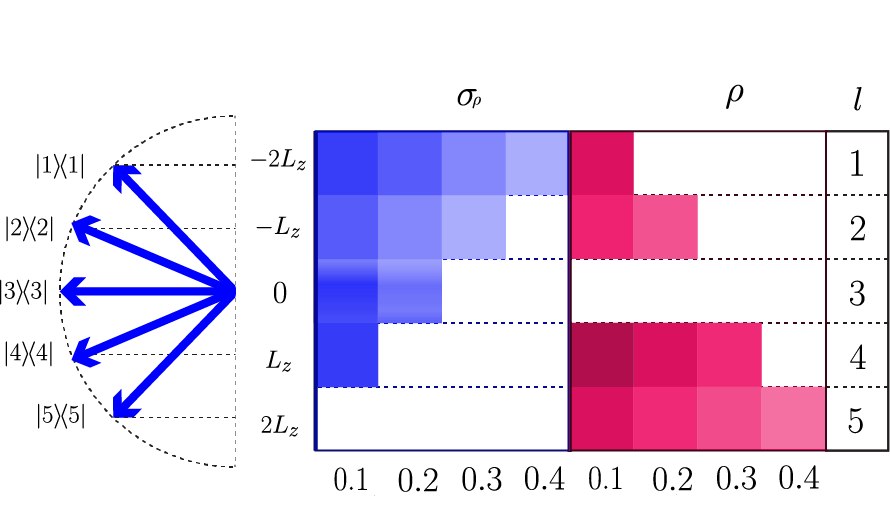}
    \caption{A 5-level system with internal Hamiltonian $H_0(L_z)=L_z\sum_{l=1}^5 (l-3) \ketbra{l}{l}$ in state $\rho=0.1 \ketbra{1}{1}+0.2\ketbra{2}{2}+0.3\ketbra{4}{4}+0.4\ketbra{5}{5}$ has an associated passive state $\sigma_\rho=0.4 \ketbra{1}{1}+0.3\ketbra{2}{2}+0.2\ketbra{3}{3}+0.1\ketbra{4}{4}$, from which one can extract the ergotropy $W_{\textrm{max}}=1.8 L_z$ by means of some unitary operation, such as $U=\ketbra{1}{5}+\ketbra{2}{4}+\ketbra{5}{3}+\ketbra{3}{2}+\ketbra{4}{1}$. Such unitary operation is not unique, since an arbitrary relative phase can be introduced for each term $\ketbra{i}{j}$.}
    \label{fig:passive-5}
\end{figure}

\noindent
A time-dependent field $V(t)$ is used to reversibly extract energy from such a battery via unitary evolution generated by $H_0+V(t)$. Given that the battery is found in some initial state described by the density operator $\rho$, the time evolution of the system is obtained from the von Neumann equation ($\hbar=1$)
\begin{equation}
    \label{eq:unitary_charging}
    \dot{\rho}(t) = -i[H_0+V(t),\rho(t)],
\end{equation}
with $\rho(0)=\rho$, and where the left-hand side represents the time derivative of $\rho(t)$. A solution of eq. \eqref{eq:unitary_charging} is given by $\rho(t)=U(t)\rho U^\dagger(t)$, where the unitary operator $U(t)$ is obtained as the time-ordered exponential of the generator $H_0+V(t)$, which can, in principle, correspond to any unitary transformation on the battery's Hilbert space $\mathcal{H}$,
\begin{equation}
    \label{eq:unitary}
    U(t)=\mathcal{T}\big\{\exp\big[-i\int_0^t ds \big(H_0+V(s)\big)\big]\big\}.
\end{equation}
{\fontfamily{phv}\selectfont \textbf{Ergotropy}}
\\
\vspace{-0.1cm}\\
\noindent
The work extracted after some time $\tau$ by such a procedure is equal to
\begin{align}
    \label{eq:work}
    W &=\tr[\rho H_0]-\tr[\rho(\tau) H_0]
    \\
    &=\tr[\rho H_0]-\tr[U(\tau)\rho U^\dagger(\tau) H_0].
\end{align}
Since we are interested in reversible work extraction, we look for the maximal amount of extractable work, known as \emph{ergotropy} \cite{Allahverdyan2004},  optimizing $W$ over all unitary operations,
\begin{equation}
    \label{eq:ergotropy}
    W_{\textrm{max}}:=\tr[\rho H_0]-\min_{U\in SU(d)}\big\{\tr[U\rho U^\dagger H_0]\big\}.
\end{equation}
When no work can be extracted from a state $\sigma$, such state is said to be \emph{passive} \cite{Pusz1978,Lenard1978,Alicki2013}. Accordingly, a state $\sigma$ is passive if $\tr[\sigma H_0]\leq\tr[U\sigma U^\dagger H_0]$ for all unitaries $U$, or, equivalently, if and only if $\sigma=\sum_{j=1}^d s_j \ketbra{j}{j}$ for $s_{j+1}\leq s_j$, \emph{i.e.} it commutes with the internal Hamiltonian $H_0$ and has non-increasing eigenvalues, as shown in ref. \cite{Alicki2013}. From such a definition it is easy to see that for any given state $\rho$ there is a unique passive state $\sigma_\rho$ that maximizes the extractable work
\begin{equation}
    \label{eq:ergotropy_passive}
    W_{\textrm{max}}=\tr[\rho H_0]-\tr[\sigma_\rho H_0],
\end{equation}
obtained via some unitary operation that rearranges the eigenvalues of $\rho$ in non-increasing order, as can be seen in the example given in fig. \ref{fig:passive-5}.

\vspace{0.5cm}
\noindent
{\fontfamily{phv}\selectfont \textbf{Bound on Extractable Work}}
\\
\vspace{-0.1cm}\\
\noindent
A practical and intuitive way to bound the extractable work of eq. \eqref{eq:ergotropy} is to consider a thermal state with the same entropy as the passive state $\sigma_\rho$, which also minimizes the energy with respect to $H_0$. Indeed, it has been shown that a lower bound to the erogotropy for some state $\rho$ is given by
\begin{equation}
    \label{eq:bound_ergotropy}
    W_{\textrm{max}}\leq \tr[\rho H_0]-\tr[\omega_{\bar{\beta}}H_0],
\end{equation}
where $\omega_{\beta}=\exp[-\beta H_0]/\mathcal{Z}$ is the canonical Gibbs state with inverse temperature $\beta$, and $\bar{\beta}$ is chosen such that the von Neumann entropy $S(\rho)=-\tr[\rho \log\rho]$ of $\rho$ is equal to that of $\omega_{\bar{\beta}}$  \cite{Alicki2013}.

\begin{figure}
    \centering
    \includegraphics[width = 0.35\textwidth]{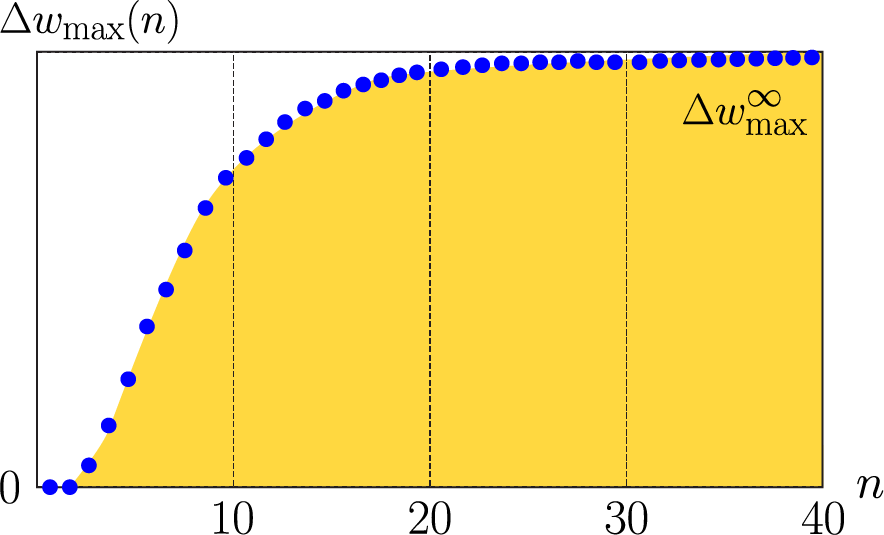}
    \caption{Let us consider an ensemble of $n$ copies of the same passive state $\sigma$. As discussed in this section, there is a non-trivial amount of work per copy $\Delta w_{\textrm{max}}(n)$ that we can extract from such an ensemble by means of entangling operations (at least 2-body operations), given by $\frac{1}{n}\{\tr[(\otimes^n\sigma-\tilde{\sigma}) H_0^{(n)}]\}$, where $\tilde{\sigma}$ is the passive state for $\otimes^n \sigma$. This figure represents this additional available energy per copy for a three-level system with energy levels $\{0,0.579,1\}$ and passive state $\sigma$ with eigenvalues $\{0.538,0.237,0.224\}$ \cite{Alicki2013}. In particular, the maximal amount of extractable energy $\Delta w^\infty_{\textrm{max}}$ is obtained in the limit of $n\to\infty$.}
    \label{fig:energy-copy}
\end{figure}

Interestingly, all thermal states are passive, and, for the case of two-level systems, all passive states are thermal, since one can always define a (positive or negative) temperature using the equation $(1-p)/p=\exp(-\beta \Delta E)$.
However, for systems with dimension greater than two, the Gibbs state $\omega_{\bar{\beta}}$ is in general different from the passive state $\sigma_\rho$ associated to $\rho$. 
Even more interestingly, the product of two or more copies of a passive state $\sigma_\rho$ is not necessarily the passive state of the copies of $\rho$, \emph{i.e.} $\otimes^n \sigma_\rho \neq \sigma_{\otimes^n \rho}$ \cite{Alicki2013}, which leads to the definition of completely passive states, as those whose $n$-copy ensemble are still passive for any $n$. It has been shown that a state is completely passive if and only if it is a thermal state \cite{Pusz1978,Lenard1978}, an observation that can be used to beat the bound given in eq. \eqref{eq:ergotropy_passive} for many copies of the same battery
by means of entangling operations \cite{Alicki2013}\\
\newline
\noindent
{\fontfamily{phv}\selectfont \textbf{Optimal Work Extraction \\ from an Ensemble of Batteries}}
\\
\vspace{-0.1cm}\\
\noindent
Consider a battery given by an ensemble of $n$ copies of the same $d$-dimensional unit cell defined by eq. \eqref{eq:internal_hamiltonian}. This new battery has an associated Hamiltonian $H_0^{(n)}$ given by the sum of the local internal Hamiltonians $H_{0,l}\otimes_{j\neq l}\mathbbm{1}_{j}$ of the subsystems that form the global system
\begin{equation}
    \label{eq:interal_array_hamiltonian}
    H_0^{(n)}=\sum_{l=1}^n H_{0,l},
\end{equation}
where we omit the identities to simplify the notation.
Recalling that $\otimes^n \sigma_\rho \neq \sigma_{\otimes^n \rho}$ our goal is to extract some additional work from $\otimes^n \sigma_\rho$ until a completely passive state is reached. In the limit of large $n$, the maximal amount of available work per copy of battery $w_{\textrm{max}}(n)$ in a state $\rho$ is tightly bounded as in eq. \eqref{eq:bound_ergotropy}, 
\begin{equation}
\label{eq:limit_work_available}
    \lim_{n\to\infty}w_{\textrm{max}}(n)=\tr[\rho H_0^{(1)}]-\tr[\omega_{\bar{\beta}} H_0^{(1)}],
\end{equation} 
where
\begin{equation}
    \label{eq:ergotropy_per_copy}
    w_{\textrm{max}}(n):=\frac{1}{n}\bigg(\tr[(\otimes^n\rho-\sigma_{\otimes^n\rho}) H_0^{(n)}]\bigg).
\end{equation}
The proof relies on the idea that for a large ensemble, the energy of the passive state $\sigma_{\otimes^n \rho}$ differs from that of $\otimes^n \omega_{\bar{\beta}}$ only by a small amount that tends to vanish as $n$ increases (see fig. \ref{fig:energy-copy} for an example).
\\

\noindent
{\fontfamily{phv}\selectfont \textbf{Entanglement Generation \& \\
 Power of Extracted Work}}
\\
\vspace{-0.1cm}\\
\noindent
The passive state $\sigma_{\otimes^n \rho}$ associated with any $\otimes^n\rho$ is diagonal in the eigenbasis of the local Hamiltonian of eq.~\eqref{eq:interal_array_hamiltonian}, thus it is separable. However, as we will see later in this section, in order to unitarily connect $\otimes^n\rho$ to its passive state, at least 2-body operations are required. This remark led Alicki and Fannes to conclude that, in order to reach optimal work extraction, the unit cells of such an $n$-fold battery have to be dynamically entangled. Their deduction turns out to be wrong, as proven in ref. \cite{Hovhannisyan2013}: If it is true that non-local operations (at least two-body operations) are required to beat the classical limit of eq. \eqref{eq:ergotropy_passive}, then it is always possible to reach optimal work extraction without creating any entanglement (see \cite{Giorgi2015} for quantum discord), at the expense of requiring more operations, and thus additional time. \\ 

\noindent
{\fontfamily{phv}\selectfont \textbf{Work extraction and indirect path:\\ Avoiding entanglement}}
\\
\vspace{-0.1cm}\\
\noindent
Let us consider a simple example that illustrates how to perform optimal work extraction from multiple copies of the same state without generating entanglement: A three-level system with (increasing) energy levels given by $\{E_1,E_2,E_3\}$, and two copies of some initial state $\rho=p\ketbra{2}{2}+(1-p)\ketbra{3}{3}$, with $p\in(0,1/2)$. The objective is to transform the initial state $\rho\otimes\rho=p^2\ketbra{22}{22}+p(1-p)(\ketbra{23}{23}+\ketbra{32}{32})+(1-p)^2\ketbra{33}{33}$ into a passive state $\sigma=(1-p)^2\ketbra{11}{11}+p(1-p)(\ketbra{12}{12}+0.21\ketbra{21}{21})+p^2\ketbra{13}{13}$ by means of the permutation that maps $\ket{33}\to\ket{11}$, $\dots$, $\ket{22}\to\ket{13}$. In order to avoid entanglement one can perform each swap in several steps, such as $\ket{33}\to\ket{13}$ followed $\ket{13}\to\ket{11}$, each of which keeps the state in a separable form at all times, if performed by means of controlled permutations and unitaries. 

This idea can be generalized to the case of arbitrary dimension $d$ and for any number $n$ of copies of the initial state: An $n$-body battery in an initial non-passive state $\rho=\diag(p_1,\cdots,p_{d^n})$, where $p_\alpha\geq0$ and $\sum_\alpha p_\alpha=1$. To perform optimal work extraction we need to evolve $\rho$ to the passive state $\sigma=\diag (s_1,\cdots,s_{d^n})$, where $s_{\alpha+1}\leq s_\alpha$ and $\bra{\alpha} \sigma\ket{\alpha} = s_\alpha=\Pi_{\alpha\beta}p_\beta$, for some permutation $\Pi_{\alpha\beta}$. To do so, each transposition $\alpha\leftrightarrow\beta$ that swaps $p_\alpha$ with $p_\beta$ is addressed separately by transforming $\ket{\alpha}$ to $\ket{\beta}$ (and vice versa) with a sequence of steps: First, $\ket{\alpha}=\ket{i_1^\alpha i_2^\alpha \cdots i_n^\alpha}$ is mapped to $\ket{\alpha'}=\ket{i_1^\beta i_2^\alpha \cdots i_n^\alpha}$, then to $\ket{\alpha''}=\ket{i_1^\beta i_2^\beta \cdots i_n^\alpha}$, and so on until it reaches $\ket{\beta}=\ket{i_1^\beta i_2^\beta\cdots i_n^\beta}$, after $n$ steps. Each of these steps is obtained by a unitary operator 
\begin{equation}
\label{eq:step_unitary}
    U_{\alpha\alpha'}(t)=\sum_{\mu\neq\alpha\alpha'}\ketbra{\mu}{\mu}+u_{\alpha\alpha'}(t),
\end{equation}
generated by some 2-body control interaction $V_{\alpha\alpha'}(t)$, that has, in principle, the power to generate bipartite entanglement. The state $\rho(t)$ of the system at time $t$ obtained via such unitary is
\begin{equation}
    \label{eq:state_step}
    \begin{split}
    \rho(t)=&U_{\alpha\alpha'}(t)\rho U_{\alpha\alpha'}^\dagger \\
    =&(p_\alpha +p_{\alpha'})\rho_1(t)\otimes\ketbra{i_2^\alpha\cdots i_n^\alpha}{i_2^\alpha\cdots i_n^\alpha}\\
    &+\sum_{\mu\neq\alpha\alpha'}p_\mu\ketbra{\mu}{\mu},
    \end{split}
\end{equation}
with $\rho_1$ being itself a state. The overall state $\rho(t)$ is thus separable at every step of the procedure, and after $2n-1$ total steps the target final state $\sigma$ is reached.

Of all the possible unitary cycles that connect the state $\otimes^n\rho$ to its passive state $\sigma_{\otimes^n\rho}$, those that preserve the system in a separable state are inevitably slower than those that generate entanglement. Accordingly, the authors of ref. \cite{Hovhannisyan2013} indicate a relation between the rate of entanglement generation and the power of work extraction -- defined as the ratio between extracted work and time required for the extraction -- leaving the open problem of quantifying such a relation to successive work. This question paves the way for the study of charging and extracting power, as described in the next section. 

\section{Powerful charging}
\noindent
We now consider the task of charging quantum batteries via unitary operations. The deposited energy is then simply the opposite of the work extracted, and as long as we consider closed systems, the two tasks are essentially equivalent. \\

\noindent
{\fontfamily{phv}\selectfont \textbf{Bounds on minimal time of evolution}}
\\
\vspace{-0.1cm}\\
\noindent
In the previous section, we saw how optimal work extraction can be performed with a sequence of unitary operations, and how the ensemble of batteries can be kept in a separable state at the expense of adding extra steps to such a sequence. If each of those operations could be performed instantly, the total number of steps would not influence the power of work extraction. However, in practice, each of those unitary operations requires a finite amount of time to be performed, which follows from the fact that Hamiltonians have finite magnitude (\emph{i.e.} they are bounded operators).
There holds a fundamental bound on the minimum time necessary to perform a unitary evolution, known as the quantum speed limit\footnote{See ref. \cite{Deffner2013} for an extended review on quantum speed limits and their applications.} (QSL), which provides an operational interpretation of the time-energy uncertainty relation $\Delta t \Delta E \geq \hbar$ \cite{Deffner2017}. 

The minimum time required to evolve some pure state $\ket{\psi}$ to another pure state $\ket{\phi}$ by means of a unitary operator $U(t)$ generated by time-dependent Hamiltonian $H(t)=H_0+V(t)$ is bounded by
\begin{equation}
    \label{eq:QSL}
    T(\ket{\psi},\ket{\phi})=\hbar \frac{\arccos|\!\braket{\psi|\phi}\!|}{\min \{E,\Delta E \}},
\end{equation}
where the numerator measures the distance\footnote{The unique distance on the space of pure states that is invariant under unitary operations is the Fubini-Study distance, given by the angle between the two considered states \cite{Bengtsson2008}.} between the states, while $E$ and $\Delta E$ correspond to the time-averaged energy (relative to the ground state) and standard deviation of the Hamiltonian $H(t)$ \cite{Deffner2013}.

\vspace{0.5cm}
\noindent
{\fontfamily{phv}\selectfont \textbf{Average and instantaneous power}}
\\
\vspace{-0.1cm}\\
\noindent 
When we address the problem of charging quantum batteries, an interesting task is to understand how to deposit energy as quickly as possible, \emph{i.e.} how to maximize average or instantaneous power.
The average power $P$ of some unitary charging between $\rho$ and $\rho(T)=U(T)\rho U^\dagger(T)$ is simply given by the ratio between the energy deposited on the battery during the procedure and the time required to perform the unitary operation, 
\begin{equation}
    \label{eq:average_power}
    \langle P\rangle=\frac{W}{T},
\end{equation}
remembering that from now on $W$ has the opposite sign with respect to that of eq. \eqref{eq:work}.
Similarly, the instantaneous power $P(t)$ at some time $t$ is given by the time derivative of the energy deposited at time $t$ along the unitary charging,
\begin{equation}
    \label{eq:istantaneous_power}
    P(t) =  \frac{d}{dt}W = \frac{d}{dt} \bigg\{ \tr[\rho(t)H_0]-\tr[\rho H_0] \bigg\},
\end{equation}
which becomes $P(t)=-i\tr\{[H_0+V(t),\rho(t)]H_0\}$, using the von Neumann equation. 

In order to address this optimization problem a constraint on the driving Hamiltonian $H+V(t)$ has to be considered, to prevent one from increasing the average power by investing more energy into the driving Hamiltonian, as to allow for fair comparison between different charging procedures. The constraint can be of the form 
\begin{equation}
    \lVert H+V(t)\rVert \leq E_{\textrm{max}},
\end{equation}
for some norm such as trace or operator norms, and some energy $E_{\textrm{max}}>0$. This constraint is operationally equivalent to limiting the energy at our disposal in order to perform the charging procedure.
Starting from those considerations, it is possible to show that entangling operations are more powerful than local ones, yielding an advantage that scales up to linearly with the number $n$ of unit cells  \cite{Binder2015,Campaioli2017}. \\

\begin{figure}
    \centering
    \includegraphics[width = 0.5\textwidth]{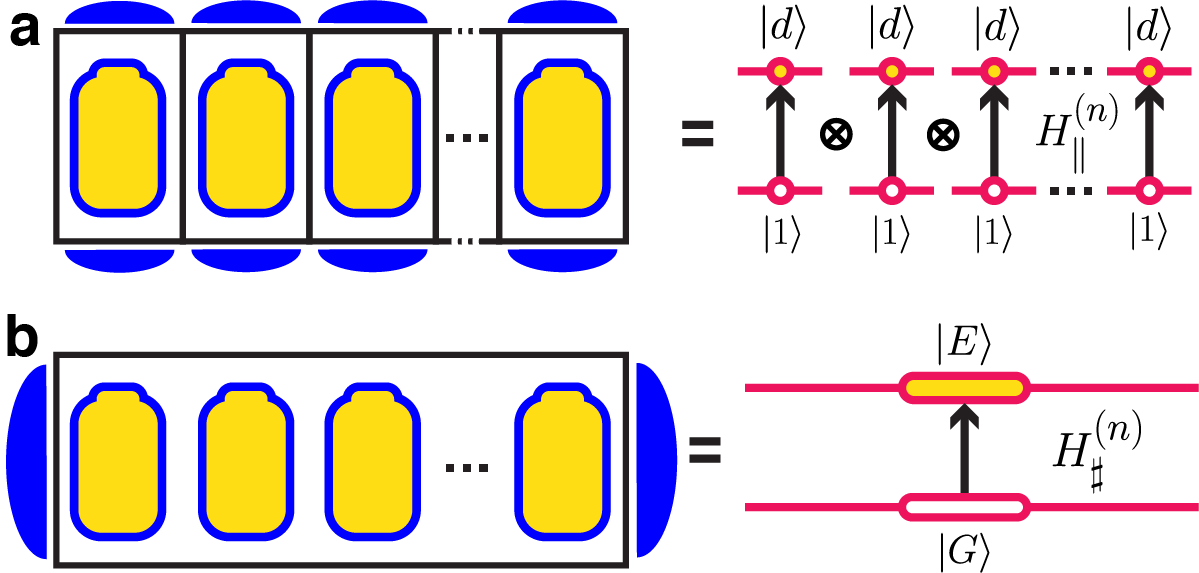}
    \caption{Local ({\fontfamily{phv}\selectfont \textbf{a}}) and global ({\fontfamily{phv}\selectfont \textbf{b}}) charging procedure are here schematically represented. The optimal local driving couples each individual ground state $\ket{1}_i$ to its respective excited state $\ket{d}_i$, while the optimal global driving couples the collective ground state $\ket{G}=\otimes^n\ket{1}$ to the collective excited one $\ket{E}=\otimes^n\ket{d}$.}
    \label{fig:parallel_collective}
\end{figure}

\noindent
{\fontfamily{phv}\selectfont \textbf{Optimal charging of an\\
array of batteries}}
\\
\vspace{-0.1cm}\\
\noindent
We now consider a battery given by $n$ copies of a $d$-dimensional unit cell, whose internal Hamiltonian is given by eq. \eqref{eq:interal_array_hamiltonian}. Assuming that the energetic structure of the individual Hamiltonians $H_l$ is the same for each copy, the highest and lowest energy states are now $\ket{G}:=\otimes^n\ket{1}$ and $\ket{E}:=\otimes^n\ket{d}$, respectively. The energy deposited onto the battery after an evolution from $\ket{G}$ to $\ket{E}$ is equal to $W(n)=n(\varepsilon_d-\varepsilon_1)$.
We are going to compare the charging power of 
\begin{align}
    \label{eq:parallel}
    & H_\|^{(n)}=\alpha_\| \sum_{l=1}^n (\ketbra{1}{d}_l+h.c.)\otimes_{j\neq l}\mathbbm{1}_{j},\\
    \label{eq:collective}
    & H_\sharp^{(n)}=\alpha_\sharp (\ketbra{E}{G}+h.c.),
\end{align}
which provide the optimal local (parallel) and global (collective) driving, respectively, when we require any Hamiltonian $H$ to satisfy the constraint\footnote{The operator norm  $\lVert \cdot \rVert_{\textrm{op}}$ of an operator $H$ is equal to its largerst singular value; if the operator $H$ is Hermitian (and any Hamiltonian is) then the operator norm is equal to the largest eigenvalue.} $\lVert H \rVert_{\textrm{op}} = E_{\textrm{max}}$, for some energy $E_{\textrm{max}}>0$ \cite{Binder2015}. We thus obtain $\alpha_\|=E_{\textrm{max}}/n$ and $\alpha_\sharp=E_{\textrm{max}}$, which allows us to express the time required to perform the two different procedures as
\begin{equation}
\label{eq:extensive_advantage}
     T_\| = n\frac{\pi}{2}\frac{1}{E_{\textrm{max}}}, \;\; T_\sharp = \frac{\pi}{2}\frac{1}{E_{\textrm{max}}}.
\end{equation}
If we calculate the average power according to eq. \eqref{eq:average_power} we obtain $P_\sharp = n P_\|$; the power of the entangling operation is $n$ times larger than that of local ones. These two charging procedures are schematically represented in fig. \ref{fig:parallel_collective}. 
Such advantage can be interpreted geometrically: While the collective Hamiltonian of eq. \eqref{eq:collective} drives the initial state along the shortest path, through the space of entangled states, the local Hamiltonian generates a longer orbit, that in return keeps the state separable for all times \cite{Binder2015}.

\vspace{0.5cm}
\noindent
{\fontfamily{phv}\selectfont \textbf{Quantum Advantage}}
\\
\vspace{-0.1cm}\\
\noindent
Consider the task of charging an $n$-body battery between some product state $\otimes^n \rho$, to some other product state $\otimes^n \sigma$ by means of cyclic unitary operations. 
We can introduce a simple parameter to quantify the advantage of using entangling operations over local ones, given by the power ratio 
\begin{equation}
    \label{eq:quantum_advantage}
    \Gamma:=\frac{\langle P \rangle}{\langle P_\| \rangle} = \frac{T_\|}{T},
\end{equation}
where $\langle P \rangle$ is the power of the considered charging procedure, while $\langle P_\| \rangle$ is the charging power obtained with the best local driving \cite{Campbell2017}. Since the energy $W(n)$ deposited onto the battery depends only on the choice of initial and final states, $\Gamma$ can be simply expressed as the ratio between the optimal parallel time $T_\|$ and the time $T$ required to perform the charging. 

The parameter $\Gamma$, called the quantum advantage, can be greater than unity only when entangling operations are considered \cite{Campaioli2017}. 
Upper bounds for $\Gamma$ are obtained using the quantum speed limit, as shown in ref. \cite{Campaioli2017}, and depend on the chosen constraint on the driving Hamiltonian $H$. In particular, imposing a constraint on the standard deviation of the driving Hamiltonian yields an advantage that scales with $\sqrt{n}$, whereas a linear scaling results from constraining the average energy of $H$ or its operator norm. 

\vspace{0.5cm}
\noindent
{\fontfamily{phv}\selectfont \textbf{Correlations and Optimal Power}}
\\
\vspace{-0.1cm}\\
\noindent
An interesting result obtained by Campaioli et al.~\cite{Campaioli2017} sheds light on the nature of the quantum advantage and its dependence on quantum correlations, such as entanglement. The authors provide an example where an initial, highly mixed state $\otimes^n\rho$ of $n$-copies of a 2-level system evolves to some final state $\otimes^n \sigma$, where $\rho=\exp[-\epsilon H_0]/{\rm tr}\{\exp[-\epsilon H_0]\}$, and $\sigma=\exp[\epsilon H_0]/{\rm tr}\{\exp[\epsilon H_0]\}$ are thermal states with inverse temperature $\pm\epsilon$, and where $H_0\propto\sigma_z$. This system can be driven with the collective Hamiltonian in eq. \eqref{eq:collective} in order to achieve a quantum advantage $\Gamma=n$ over the optimal local charging, for any value of $\epsilon>0$. 

\begin{figure}
    \centering
    \includegraphics[width = 0.5\textwidth]{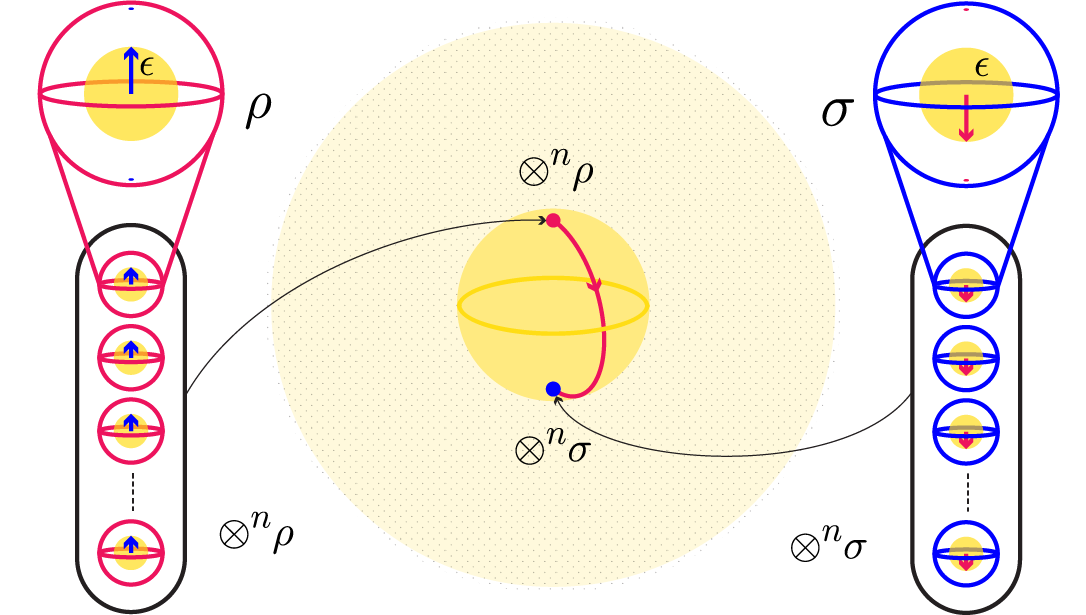}
    \caption{In the example described by Campaioli et al., a product state $\otimes^n\rho$ of $n$ copies of the same highly mixed state $\rho$ of a 2-level system is unitarily evolved into $\otimes^n\sigma$, where $\rho=\exp[-\epsilon \sigma_z]/\mathcal{Z}$, and $\sigma=\exp[\epsilon \sigma_z]/\mathcal{Z}$. For large $n$, if the inverse temperature $\epsilon$ is chosen to be small enough, the state of the copies lies in a region of the space of states that contains only separable states, called separable ball. Since unitary evolution cannot change the purity of a state, any driving that maps such $\otimes^n\rho$ to $\otimes^n\sigma$ preserves the system in a separable state at all times. The authors of ref. \cite{Campbell2017} show that the global Hamiltonian given in eq. \eqref{eq:collective} can lead to a quantum advantage $\Gamma=n$.}
    \label{fig:sep_ball}
\end{figure}

However, as $\epsilon$ decreases the collective state $\otimes^n\rho$ gets closer and closer to the separable ball, a spherically symmetric region of the space of states, centered around the maximally mixed state, that contains only separable states. Since the distance from the maximally mixed state is invariant under unitary evolution, the final state $\otimes^n\sigma$, along with every other state through which the evolution proceeds, will also be in the separable ball if the initial state $\otimes^n\rho$ is (see fig. \ref{fig:sep_ball}). The authors show that for any $n$ there is an $\epsilon>0$ small enough that the evolution can be performed entirely within the separable ball, yet still with a linear advantage in $n$ with respect to the optimal local procedure. This leads to the conclusion that entanglement \emph{per se} is not required for a quantum advantage in the charging power of quantum batteries. 

The role of other forms of quantum correlations, such as quantum discord \cite{Ollivier2001}, has not been investigated yet; however, as we show in the next section, this example hints that a possible resource is given by the order $k$ of non-local $k$-body interactions that are available for the charging task.

\vspace{0.5cm}
\noindent
{\fontfamily{phv}\selectfont \textbf{Feasible Charging Scenarios}}
\\
\vspace{-0.1cm}\\
\noindent
The advantage that can be obtained using entangling operations such as that in eq. \eqref{eq:collective} might be practically hard to obtain. The reason is that such collective interactions are given by highly non-local terms\footnote{For example, in the case of 2-level systems such interactions have non-trivial components proportional to products of $n$ Pauli operators $\sigma_i^{(1)}\otimes\cdots\sigma_j^{(n)}$, with $i=1,2,3$.}.
As these interactions are naturally rare or hard to engineer, it is useful to understand how to maximize the quantum advantage when we are limited to $k$-body interactions, with $2\leq k<n$. 

\begin{figure}
    \centering
    \includegraphics[width = 0.35\textwidth]{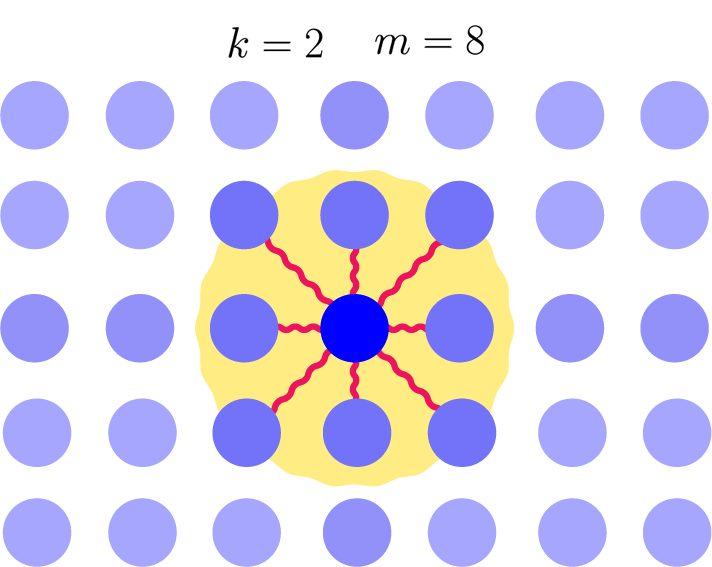}
    \caption{Let us consider a battery given by $n$ unit cells arrange in a 2-d lattice, where each unit cell is coupled with its neighbours via 2-body interactions ($k=2$). If the reach of these interactions is limited to the nearest neighbours, each unit cell interacts with up to 8 others ($m=8$). Under this conditions the achievable quantum advantage $\Gamma$ is bounded as in eq. \eqref{eq:QA_feasible}.}
    \label{fig:realisitc}
\end{figure}

It is possible to derive a bound on $\Gamma$ under the assumption that the operator norm of any driving Hamiltonian is bounded by some positive quantity $E_{\textrm{max}}$, the interaction order is at most $2\leq k<n$, and the participation number, \emph{i.e.} the maximum number of interactions that each unit cell can take part in, is at most $m>1$, obtaining 
\begin{equation}
    \label{eq:QA_feasible}
    \Gamma<\gamma [k^2(m-1)+k],
\end{equation}
where $\gamma$ is a constant factor that does scale with the number $n$ of cells (see fig. \ref{fig:realisitc} for an example) \cite{Campaioli2017}. Such a result is obtained by performing a Trotterization of the unitary evolution, decomposed in terms of a set of $k(m-1)+1$ circuits, each of which is given by piece-wise time-independent Hamiltonians that have up to $k$-body interactions.

On one hand, this result rules out the possibility of increasing the number of cells in the battery in order to enhance its charging power. On the other, it clearly shows the importance of the order $k$ of the interactions as an effective resource in these charging schemes.

\vspace{0.5cm}
\noindent
{\fontfamily{phv}\selectfont \textbf{Cycle Precision}}
\\
\vspace{-0.1cm}\\
\noindent
Earlier in this section we discussed the importance of time-energy uncertainty relations, and how they set a bound on the minimum time required to perform the unitary cycles that we considered to charge quantum batteries. As large power corresponds to short times, short times lead to large work fluctuations, which, in turn, lead to inaccuracy. Such trade-offs between a cycle's precision and power have to be carefully addressed, in order to avoid the likely scenario of depositing a hazardous amount of energy \cite{Friis2017a}. Charging precision can be explored minimizing either the standard deviation of the final energy, or the fluctuations during the charging process \cite{Friis2017a}. 

This problem has been addressed by Friis and Huber who considered a model given by a number of harmonic oscillators in the canonical Gibbs state $\omega_\beta$, where $\beta$ is the inverse temperature of the battery, which are then charged by means of unitary operations. 
The authors suggest the use of Gaussian unitary operations, \emph{i.e.} those unitary operations that preserve the system in the space of Gaussian states, motivated by the fact that such operations are of accessible experimental realization. Gaussian operation also represent a trade-off between performance and feasibility: The authors are able to identify pure single-mode squeezing as the least-favourable operations, when one wishes to obtain precise charging of single-mode batteries, while they indicate combinations of squeezing and displacements as the most accurate and accessible operations  \cite{Friis2017a}. 

\section{Possible Implementations}
\noindent
The study of quantum batteries is fundamentally driven by the ambition of realizing devices of atomic and molecular size, which could gain a considerable advantage over their macroscopic counterparts for some particular tasks, so, as theoretical advancements lead the way, we must also consider experimental realizations: Fundamental milestones, such as the implementation of optimal extraction protocols, or that of an extensive quantum advantage, would be of exceptional impact. 

Here we summarize some of the possible experimental scenarios that authors have indicated, such as spin-chains \cite{Le2018}, nanofabricated quantum dots coupled to cavities, and superconducting qubits \cite{Ferraro2018b}.\\

\noindent
{\fontfamily{phv}\selectfont \textbf{Cavity Assisted Charging}}
\\
\vspace{-0.1cm}\\
\noindent
The collective terms required to maximize the charging power can be recovered by coupling the batteries to an electromagnetic field, where collective coherences lead to superradiance \cite{Campaioli2017,Ferraro2018b}. In particular, Ferraro et al. use a Dicke model to powerfully charge an array of 2-level systems coupled with a quantized single-mode electromagnetic field \cite{Ferraro2018b}.

\begin{figure}
    \centering
    \includegraphics[width = 0.48\textwidth]{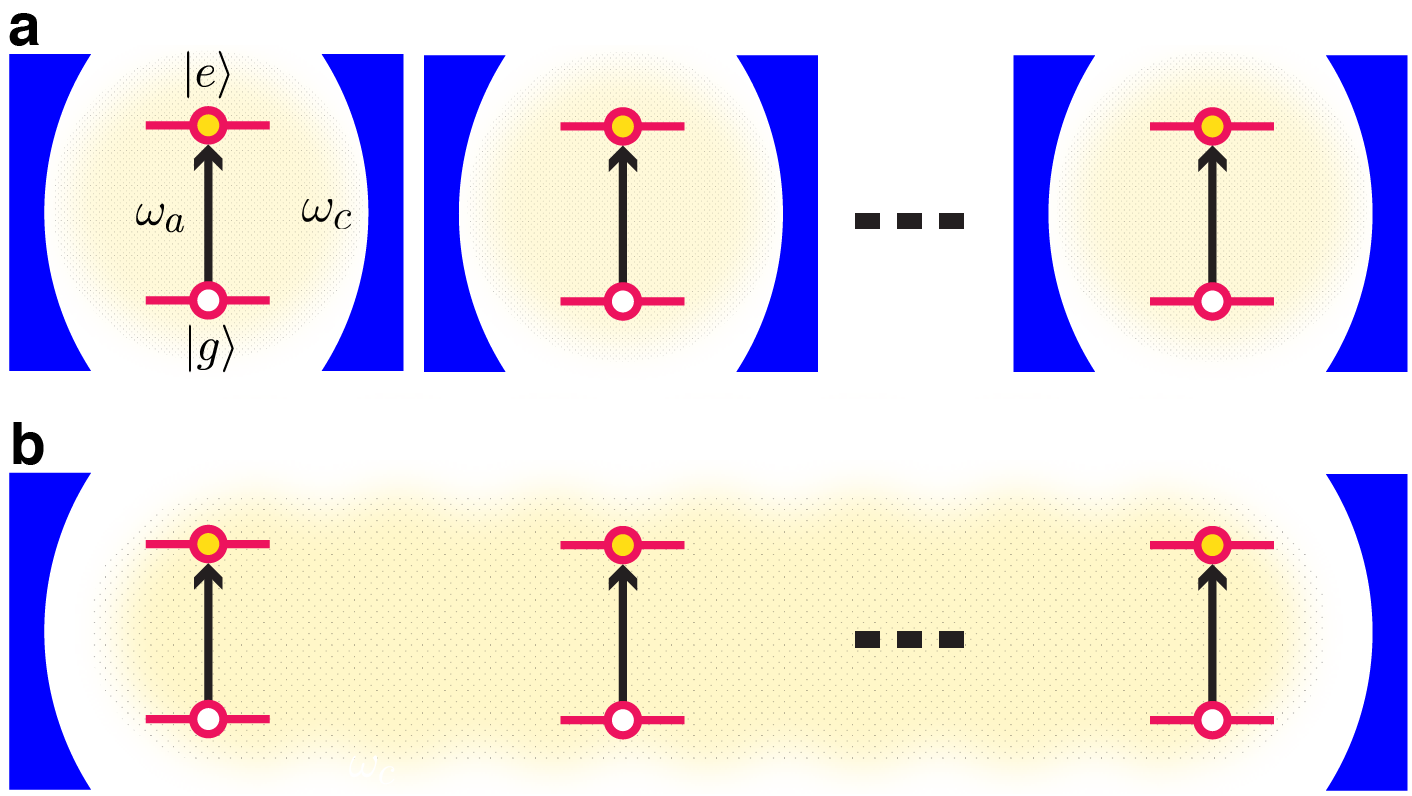}
    \caption{Ferraro et al. propose a Dicke model for a cavity-assisted quantum battery that can be charged both locally ({\fontfamily{phv}\selectfont \textbf{a}}) and globally ({\fontfamily{phv}\selectfont \textbf{b}}), as represented by this scheme. The cavity's single mode frequency is given by $\omega_c$, while the energy-splitting of each 2-level system is characterized by $\omega_a$. This model can be used to obtain effective $n$-body interactions between the $n$ unit cells, and achieve a quantum advantage $\Gamma\propto\sqrt{n}$.}
    \label{fig:dicke}
\end{figure}

The model considered -- schematically represented in fig. \ref{fig:dicke} -- is given by the time-dependent Dicke Hamiltonian
\begin{equation}
\label{eq:dicke}
    H^{(n)}=\hbar \omega_c \hat{a}^\dagger\hat{a}+\omega_a \hat{J}_z+2\omega_c\lambda_t\hat{J}_x(\hat{a}+\hat{a}^\dagger),
\end{equation}
where $\hat{a},\hat{a}^\dagger$ are the creation and annihilation operator for the single-mode cavity with frequency $\omega_c$, while
\begin{equation}
    \hat{J}_i=\frac{\hbar}{2}\sum_{l=1}^n\sigma_i^{(l)}, 
\end{equation}
are the components of the collective spin operators expressed in terms of Pauli operators $\sigma_i^{(l)}$ of the $l$-th 2-level system. The energy splitting of each 2-level system is given by $\hbar \omega_a$, which is tuned in order to reach resonance with the cavity, \emph{i.e.} $\omega_a=\omega_c$. Control is achieved by adjusting the strength of the coupling $\lambda_t$ between cavity and array.

First, the system is initialized in state 
\begin{equation}
    \label{eq:initial_dicke}
    \ket{\psi^{(n)}(0)}=\ket{n}\otimes\ket{G},
\end{equation}
where $\ket{G}=\otimes^n\ket{g}$, and 
where $\ket{n}$ is the Fock state of the cavity with $n$ photons in its single mode, and where $\ket{g}$ is the ground state of the 2-level system.
Charging is performed by turning on the coupling $\lambda_t$ to some fixed value $\bar{\lambda}$, before switching it off at a later time $\tau_c$. 
Note that, in this case, while battery and cavity undergo the unitary evolution the battery alone evolves according to a non-unitary map. In particular, the energy $W(\tau_c)$ stored at time $\tau_c$ is calculated with respect to the array of two-level systems (\emph{i.e.} the battery), and thus associated with $\omega_c \hat{J}_z$, rather than the whole internal Hamiltonian $H_0^{(n)} = \omega_c \hat{J}_z+\hbar\omega_c \hat{a}^\dagger\hat{a}$. 

The authors compare the maximum charging power that can be obtained in the collective case with the one that can be obtained via parallel charging, where the latter corresponds to a collection of $n$ independent Hamiltonians $H^{(1)}$. They show that, in the limit of large $n$, the best charging power of the collective approach is up to $\sqrt{n}$ times larger than that of the parallel approach
\begin{equation}
\label{eq:power_ratio}
    \frac{\textrm{max} [\langle P_\sharp \rangle]}{\textrm{max} [\langle P_\| \rangle]}\leq \sqrt{n},
\end{equation}
modulo some constant factor that does not depend on the number $n$ of 2-level system considered in the protocol \cite{Ferraro2018b}.

In the same work, Ferraro and colleagues discuss the tasks of storage and discharging, and the feasibility of the solid-state Dicke model discribed in their work, which can be realized by means of superconducting qubits or nanofabricated semiconductor quantum dots \cite{Ferraro2018b}. We now review another possible implementation of quantum batteries, proposed by Le et al., which consists in a spin-chain model characterized by many-body interactions \cite{Le2018}.

\vspace{0.5cm}
\noindent
{\fontfamily{phv}\selectfont \textbf{Spin-chain Battery}}
\\
\vspace{-0.1cm}\\
\noindent
The common denominator of the batteries described so far in this chapter is the local nature of the internal Hamiltonian given by eq. \eqref{eq:interal_array_hamiltonian}. As an alternative to such models one can consider a system composed of $n$ unit cells whose internal Hamiltonian contains $k$-body interactions. 

Le et al. specifically consider a one-dimensional Heisenberg spin-chain (2-level systems) characterized by 2-body interactions ($k=2$) with arbitrarily long range, which can lead to large values $m$ of the participation number. Their internal Hamiltonian $H_0^{(n)}=H_B+H_g$ is given by two terms,
\begin{align}
\label{eq:spin-chain_field}
    H_B=&B\sum_{i=1}^n\sigma_z^{(i)},\\
    \label{eq:spin-chain_Hamiltonian}
    \begin{split}
    H_g=&-\sum_{i<j}g_{ij}[\sigma_z^{(i)}\otimes\sigma_z^{(j)}+\\
    &+\alpha(\sigma_x^{(i)}\otimes\sigma_x^{(j)}+\sigma_y^{(i)}\otimes\sigma_y^{(j)})],
\end{split}
\end{align}
where $g_{ij}$ is the interaction strength between different spins, while $\alpha$ can be tuned to recover Ising ($\alpha=0)$, XXZ ($0<\alpha<1$), and XXX ($\alpha=1)$ Heisenberg models, respectively \cite{Le2018}.
The system is then charged using an external field $V=\omega\sum_{i}\sigma_x^{(i)}$, while turning off $H_B$, in order to obtain the driving Hamiltonian $H^{(n)}=H_g+V$, which generates a time-independent unitary evolution $U_t=\exp[-i H t]$.

The remarkable difference between this model and the others reviewed so far is that the eigenstates of the internal Hamiltonian $H_0^{(n)}$ can be entangled, if the coupling strength $g_{ij}$ is non-vanishing, due to the presence of 2-body interactions between the spins. 
For this reason, rather than following a path from an initial to a final separable state, the authors study a quantum advantage $\Gamma=\langle P \rangle/\langle P_{\textrm{ind}} \rangle$ that is given by the ratio between the power $\langle P \rangle$ of some charging strategy and that $\langle P_{\textrm{ind}} \rangle$ of the best independent driving ($g_{ij}=0$). In particular, in this case the energy can be \emph{stored} in the interactions between the different spins. 

The findings of Le and coauthors shed some light on the importance of correlations and many-body interaction for the charging power:
first of all, they show that the isotropic coupling of the XXX Heisenberg model, \emph{i.e.} $\alpha=1$, leads to completely independent charging of each spin, thus to no advantage. Conversely, the full anisotropy of the XXZ model, \emph{i.e.} $\alpha=-1$, leads to much higher power compared to the independent case. 

Another important result is obtained when authors study the case of strong coupling $g\gg\omega$ in order to recover effective $n$-body interactions. They find that the strength of such effective interactions in their model necessarily decreases with their ability to produce them, since they are only valid in a perturbative regime. As a result, the charging power of the battery is actually worse than that of the independent charging, and vanishes in the limit of large $n$. 

\begin{figure}
    \centering
    \includegraphics[width = 0.5\textwidth]{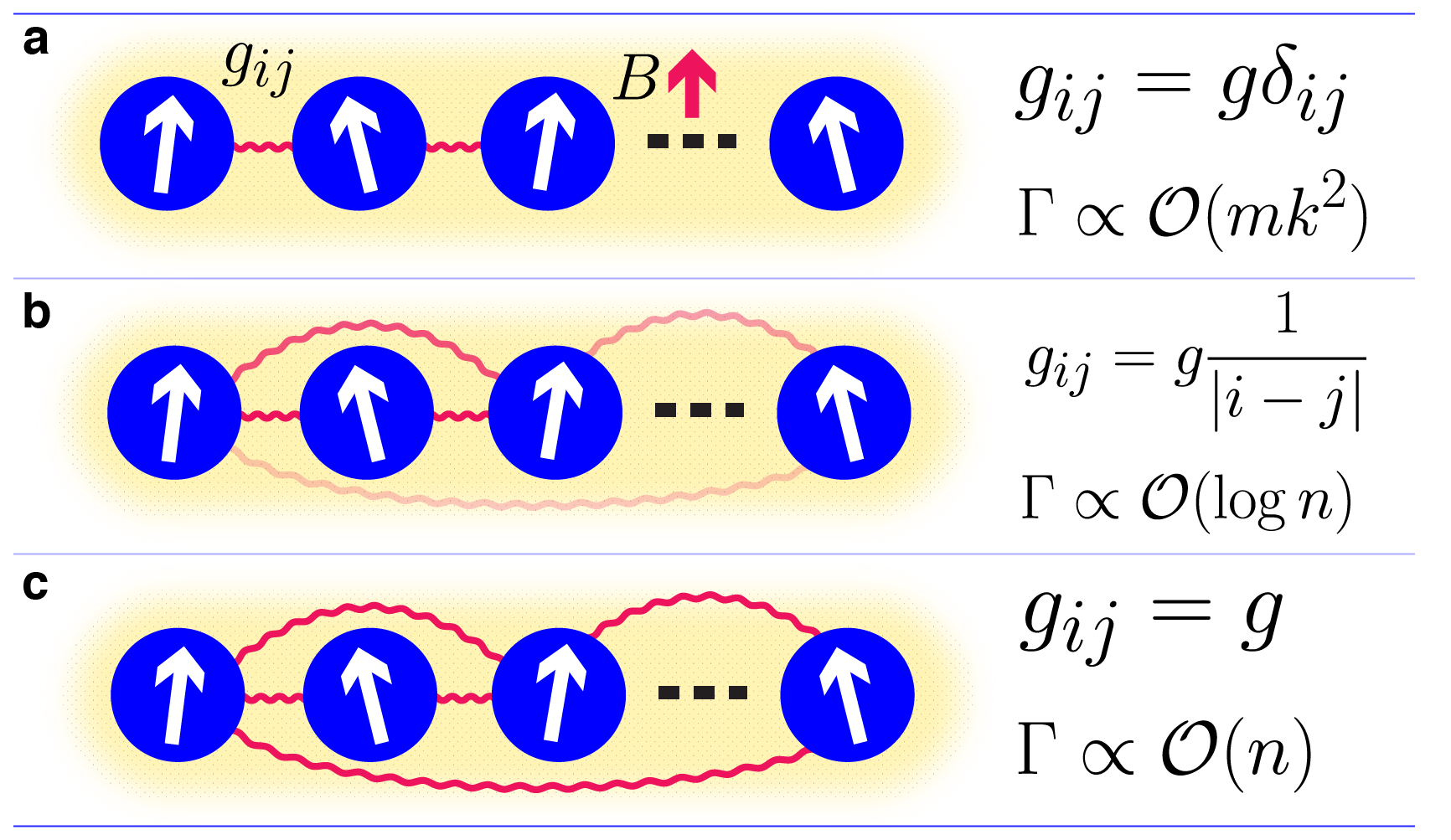}
    \caption{\noindent 
    Le et al. consider a many-body spin-chain model to obtain a quantum battery in ref. \cite{Le2018}, as described in eqs. \eqref{eq:spin-chain_Hamiltonian} and \eqref{eq:spin-chain_field}. They study the dependence of the achievable quantum advantage on the interaction range in the weak coupling
    regime ($\sum_{i<j}g_{ij}\ll n \omega$), showing that nearest-neighbour interactions ({\fontfamily{phv}\selectfont \textbf{a}}) lead to an advantage that does not depend on the number $n$ of unit cells, long-range interactions ({\fontfamily{phv}\selectfont \textbf{b}}) $g_{ij}\propto 1/|i-j|$ lead to $\Gamma\propto\mathcal{O}(\log n)$, while uniform interactions ({\fontfamily{phv}\selectfont \textbf{c}}) $g_{ij}=g$ lead to an extensive advantage. Their findings are in agreement with eq. \eqref{eq:QA_feasible}, when order of the interaction $k$ and participation number $m$ are considered.}
    \label{fig:spin_chain}
\end{figure}

An interesting quantum advantage for the charging power becomes evident in the weak interaction case, \emph{i.e.} when $\sum_{i<j}g_{ij}\ll n \omega$. 
In particular, if the spin-spin interactions have finite range (or if the coupling strength decays more rapidly as their distance increase), such as for nearest neighbour interactions (see fig. \ref{fig:spin_chain} {\fontfamily{phv}\selectfont \textbf{a}}), the power is enhanced only by a factor that is constant in $n$. When the coupling strength decays as $1/|i-j|$, where $i$ and $j$ are ordered spins $i$ and $j$ along the chain, the power grows super-extensively, and an advantage proportional to $\log n$ can be achieved (see fig. \ref{fig:spin_chain} {\fontfamily{phv}\selectfont \textbf{b}}). Finally, for uniform interaction strength the authors recover the extensive power advantage described by eq. \eqref{eq:extensive_advantage} (see fig. \ref{fig:spin_chain} {\fontfamily{phv}\selectfont \textbf{c}}). 
These results are in full agreement with eq. \eqref{eq:QA_feasible}, when the interaction is $k=2$, and when the participation number $m$ is limited by the specific range of the coupling ($m$ can be arbitraily large in the last case), which reflect the symmetry between the roles of internal ($H_0$) and driving ($H$) Hamiltonians: Here, the driving Hamiltonian is local and does not increase the correlations between different spins, while the power advantage is given by the \emph{extra} energy stored in the 2-body interactions provided by the internal Hamiltonian. 

Le's results also reinforce one of the results by Campaioli et al. about the role of correlations: The infinite-ranged interacting spin chain behaves like a global classical spin in the limit of large $n$, while providing the largest power enhancement, supporting the conclusion that quantum correlations are unnecessary for the efficient operation of such a many-body battery \cite{Le2018, Campaioli2017}. \\

\noindent
{\fontfamily{phv}\selectfont \textbf{Coupled Cooper Pair Boxes}}
\\
\vspace{-0.1cm}\\
\noindent
As mentioned earlier in this section, another implementation could be obtained by means of cooper pair boxes (CPBs) charged via gate pulses \cite{Ferraro2018b}. The motivation of using such systems comes from the fact that they have well defined charging energies, they are scalable and can be produced in large numbers, they are relatively easy to fabricate and allow for the design of local and two-body interactions.

A CPB comprises a small conducting island, typically made of aluminum, coupled to a conducting reservoir via a small Josephson junction. The junction allows for the tunneling of a Cooper pair, \emph{i.e.} a pair of electrons bonded by lattice deformation, between reservoir and island. Restricted to a pair of adjacent energy levels, such a device provides a 2-level system with logical states $\ket{0}$ and $\ket{1}$, associated to the absence/presence of an extra Cooper pair on the island, that can be used to implement 1-body and 2-body quantum gates \cite{Pashkin2009}, such as NOT and SWAPS. Due to the small capacitance of the island, the tunneling of a pair requires a significant amount of energy, which can also be adjusted by a gate voltage. Since CPBs are characterized by a well-defined charging energy they are also called \emph{charge qubits}\footnote{More precisely, CPBs are referred to as charge qubits in the large charging energy regime; when the charging energy is small, they are referred to as flux qubits.}, and could be thought of as elementary 2-level batteries, such as the one described in fig. \ref{fig:2-levle-battery}. 

Since pairs of CPBs can be coupled to each other by means of junctions between their islands \cite{Pashkin2009}, engineering two-body interactions that could be used to obtain an advantage for the charging power becomes a promising, as well as challenging task. As mentioned earlier in this chapter, such advantage would benefit of the order of the interactions $k$ (here $k=2$), resulting in a charging time that could be up to two times shorter than that of any local charging procedure, when fairly compared to the case of independent CPBs. 

Along with the practical implementations that we have discussed, there are many other problems that have to be addressed in order to bring quantum batteries to their first realizations, which we discuss in the next section.

\section{Discussion}
\label{sec:discussion}
\noindent
Until now we have considered reversible charging and extraction procedures performed via cyclic unitary operations, where initial and final states of the battery are invariant under the action of the internal Hamiltonian $H_0$: In such a case, once the driving field $V(t)$ is turned off the battery remains in the desired charged state for arbitrarily long times. However, it is sensible to think that such small and sensitive systems would interact, albeit weakly, with the surrounding environment, which could dissipate the energy stored.

In order to preserve the energy stored for reasonably long times it is fundamental to understand how to counteract the action of the environment. The latter could manifest itself by inducing decoherence, amplitude damping, and energy dissipation. Such a task could be approached with tools of quantum optimal control \cite{Wang2013,Glaser2015a,suri2017}, but nobody has addressed this problem for the specific case of quantum batteries yet. On one hand it is easy to identify the most interesting tasks, such as energy stabilization, powerful charging and optimal work extraction. Instead, modelling system-environment dynamics requires more attention, since it is not clear how quantum batteries would be realized in practice. For this reason, looking for experimental realization is just as important as optimizing charging, extraction and stability under the effect of the environment.

\bigskip


\bibliography{Chapter.bbl}

\end{document}